\documentstyle{article}

\begin{document}

\title{Random matrix models with log-singular level confinement: method of
fictitious fermions \footnote{Presented at the MINERVA Workshop on Mesoscopics, Fractals 
and Neural Networks, \newline \centerline{Eilat, Israel, March 1997}}}

\author{E. Kanzieper and V. Freilikher\vspace{0.1in} \\
The Jack and Pearl Resnick Institute of Advanced Technology,\\
Department of Physics, Bar-Ilan University, 52900 Ramat-Gan, Israel %
\vspace{0.75in}}
\date{April 17, 1997}
\maketitle

\begin{abstract}
Joint distribution function of $N$ eigenvalues of $U\left( N\right) $
invariant random-matrix ensemble can be interpreted as a probability density
to find $N$ fictitious non-interacting fermions to be confined in a
one-dimensional space. Within this picture a general formalism is developed
to study the eigenvalue correlations in non-Gaussian ensembles of large
random matrices possessing non-monotonic, log-singular level confinement. An
effective one-particle Schr\"odinger equation for wave-functions of
fictitious fermions is derived. It is shown that eigenvalue correlations are
completely determined by the Dyson's density of states and by the parameter
of the logarithmic singularity. Closed analytical expressions for the
two-point kernel in the origin, bulk, and soft-edge scaling limits are
deduced in a unified way, and novel universal correlations are predicted
near the end point of the single spectrum support.\vspace{0.2in}\\{\tt 
cond-mat/9704149}
\end{abstract}

\newpage\ 

\section{Introduction and basic relations}

Random matrices are the field-theoretical models which study the universal
features of physical phenomena resulting from the symmetry constraints only.
This is the reason why quite different physical problems get a unified
mathematical description within the framework of the random-matrix theory 
\cite{Mehta-1991}. In particular, the applicability of so-called invariant
matrix model given by the joint distribution function 
\begin{equation}
\rho _N\left( \left\{ \lambda \right\} \right) d\left\{ \lambda \right\}
=\frac 1{{\cal Z}_N}\prod_{k=1}^Nd\lambda _k\left| \lambda _k\right|
^{\alpha \beta }\exp \left\{ -\beta v\left( \lambda _k\right) \right\}
\prod_{i>j=1}^N\left| \lambda _i-\lambda _j\right| ^\beta  \label{eq.01}
\end{equation}
of $N$ eigenvalues $\left\{ \lambda \right\} $ of a large $N\times N$ random
matrix ${\bf H}$ ranges from the problem of two-dimensional gravity \cite
{Zinn-1995}, through the spectral properties of the Dirac operator in
quantum chromodynamics \cite{Verbaarschot-1994}, to the mesoscopic electron
transport in normal and superconducting quantum dots \cite
{Beenakker-1997,Atland-Zirnbauer-1996}. Here the eigenvalues $\left\{
\lambda \right\} $ belong to entire real axis, $-\infty <\left\{ \lambda
\right\} <+\infty $, and the partition function ${\cal Z}_N$ is determined
from the normalization condition $\int \rho _N\left( \left\{ \lambda
\right\} \right) d\left\{ \lambda \right\} =1$. Parameter $\beta $ in Eq. (%
\ref{eq.01}) accounts for the symmetry of the problem, $\alpha $ is a free
parameter associated with a logarithmic singularity, while $v\left( \lambda
\right) $ is a non-singular part of the confinement potential 
\begin{equation}
V\left( \lambda \right) =v\left( \lambda \right) -\alpha \log \left| \lambda
\right| .  \label{eq.02}
\end{equation}
It is implied that dimension $N$ of the matrix ${\bf H}$ is large enough, $%
N\gg 1$. In this thermodynamic limit the matrix model Eq. (\ref{eq.01})
becomes exactly solvable.

Different physics, that is behind the model introduced, deals with different
regions of spectrum that can be explored in the corresponding scaling
limits. Up to now, the most study received the random-matrix ensemble Eq. (%
\ref{eq.01}) with $U\left( N\right) $ symmetry $\left( \beta =2\right) $,
where three types of universal correlations have been established in the
origin \cite{Nishigaki-1996,Akemann-1996}, bulk \cite
{Brezin-Zee-1993,Eynard-1994,Weidenmuller-1995,FKY-1996}, and soft-edge \cite
{Airy-1997,Kanzieper-1997} scaling limits. Corresponding eigenvalue
correlations are described by the universal Bessel, sine, and \rm{G}$-$%
multicritical kernels, respectively. Various scaling limits of the model Eq.
(\ref{eq.01}) have been investigated by using different methods, so that a
unified treatment of the problem of spectral correlations in $U\left(
N\right) $ invariant ensembles is still absent. The purpose of this paper is
to present a unified approach allowing us to explore the spectral properties
of the $U\left( N\right) $ invariant matrix model Eq. (\ref{eq.01}) with
effective log-singular level confinement in an arbitrary spectrum range.

The following representation \cite{Mehta-1991} of the joint distribution
function $\rho _N\left( \left\{ \lambda \right\} \right) $ is well-known in
the random-matrix theory: 
\begin{equation}
\rho _N\left( \left\{ \lambda \right\} \right) =\left| \Psi _0\left( \lambda
_1,...,\lambda _N\right) \right| ^2,  \label{eq.03}
\end{equation}
\begin{equation}
\Psi _0\left( \lambda _1,...,\lambda _N\right) =\frac 1{\sqrt{N!}}\det
\left\| \varphi _{j-1}\left( \lambda _i\right) \right\| _{i,j=1...N}.
\label{eq.03a}
\end{equation}
As far as $\Psi _0$ takes the form of the Slater determinant, $\rho _N\left(
\left\{ \lambda \right\} \right) $ can be thought of as a probability
density to find $N$ non-interacting fictitious fermions in the quantum
states $\varphi _0,...,\varphi _{N-1}$ at the ``spatial'' points $\lambda
_1,...,\lambda _N$. The ``wave-functions'' of such fermions are uniquely
determined by the set of polynomials $P_n\left( \lambda \right) $ orthogonal
on the entire real axis with respect to the measure $d\mu \left( \lambda
\right) =\exp \left\{ -2V\left( \lambda \right) \right\} d\lambda $, 
\begin{equation}
\varphi _n\left( \lambda \right) =P_n\left( \lambda \right) \exp \left\{
-V\left( \lambda \right) \right\}  \label{eq.04a}
\end{equation}
so that the orthogonality relation 
\begin{equation}
\int_{-\infty }^{+\infty }d\lambda \varphi _n\left( \lambda \right) \varphi
_m\left( \lambda \right) =\delta _{nm}  \label{eq.04b}
\end{equation}
holds. It follows from Eq. (\ref{eq.03}) that the joint distribution
function $\rho _N\left( \left\{ \lambda \right\} \right) $ can be
represented as 
\begin{equation}
\rho _N\left( \left\{ \lambda \right\} \right) =\frac 1{N!}\det \left\|
K_N\left( \lambda _i,\lambda _j\right) \right\| _{i,j=1...N},
\label{eq.07}
\end{equation}
where $K_N\left( \lambda ,\lambda ^{\prime }\right) $ (referred to as the
``two-point kernel'') 
\begin{equation}
K_N\left( \lambda ,\lambda ^{\prime }\right) =\sum_{k=0}^{N-1}\varphi
_k\left( \lambda \right) \varphi _k\left( \lambda ^{\prime }\right) .
\label{eq.07a}
\end{equation}
is completely determined by the wave-functions $\varphi _n$. Due to an
additional constraint on the wave-functions of three successive quantum
states that results from the recurrence equation Eq. (\ref{eq.06}) below,
only the highly excited states, $\varphi _{N-1}$ and $\varphi _N$,
contribute to the two-point kernel in accordance with the
Christoffel-Darboux theorem \cite{Szego-monograph}: 
\begin{equation}
K_N\left( \lambda ,\lambda ^{\prime }\right) =c_N\frac{\varphi _N\left(
\lambda ^{\prime }\right) \varphi _{N-1}\left( \lambda \right) -\varphi
_N\left( \lambda \right) \varphi _{N-1}\left( \lambda ^{\prime }\right) }{%
\lambda ^{\prime }-\lambda }.  \label{eq.08}
\end{equation}
This formula simplifies significantly the mathematical calculations in the
thermodynamic limit $N\gg 1$. Effective Schr\"odinger equation for $\varphi
_N$, that is the cornerstone of our unified approach, will be derived in the
next Section.

\section{Effective Schr\"odinger equation}

In the particular case of the Gaussian unitary ensemble (GUE) the
wave-functions $\varphi _n\left( \lambda \right) $ are well-known. They are
eigenfunctions of a fermion confined by a parabolic potential \cite
{Mehta-1991}. For general non-Gaussian ensemble Eq. (\ref{eq.01}) the
calculation of such effective wave-functions can be done by an extension of
the Shohat's method \cite{Shohat-1930,Bonan-Clark-1986} that previously has
been used by the authors \cite{Kanzieper-1997} to treat the problem of
eigenvalue correlations in random-matrix ensembles with non-singular, strong
level confinement. This method allows us to map a three-term recurrence
equation 
\begin{equation}
\lambda P_{n-1}\left( \lambda \right) =c_nP_n\left( \lambda \right)
+c_{n-1}P_{n-2}\left( \lambda \right)  \label{eq.06}
\end{equation}
for polynomials $P_n\left( \lambda \right) $ orthogonal on the entire real
axis with respect to the measure $d\mu \left( \lambda \right) =\exp \left\{
-2V\left( \lambda \right) \right\} d\lambda ,$ 
\begin{equation}
\int_{-\infty }^{+\infty }d\mu \left( \lambda \right) P_n\left( \lambda
\right) P_m\left( \lambda \right) =\delta _{nm},  \label{eq.05}
\end{equation}
onto a second-order differential equation for corresponding fictitious
wave-functions $\varphi _n$. Coefficients $c_n$ appearing in Eq. (\ref{eq.06}%
) are uniquely determined by the measure $d\mu $.

In order to derive an effective Schr\"odinger equation, we note the
following identity 
\begin{equation}
\frac{dP_n\left( \lambda \right) }{d\lambda }=A_n\left( \lambda \right)
P_{n-1}\left( \lambda \right) -B_n\left( \lambda \right) P_n\left( \lambda
\right),  \label{eq.17}
\end{equation}
with functions $A_n\left( \lambda \right) $ and $B_n\left( \lambda \right) $
to be determined from the following consideration. Since $dP_n\left( \lambda
\right) /d\lambda $ is a polynomial of the degree $n-1$, it can be
represented \cite{Szego-monograph} through the Fourier expansion in the
terms of the kernel $Q_n\left( t,\lambda \right) =\sum_{k=0}^{n-1}P_k\left(
\lambda \right) P_k\left( t\right) $ as: 
\begin{equation}
\frac{dP_n\left( \lambda \right) }{d\lambda }=\int_{-\infty }^{+\infty }d\mu
\left( t\right) \frac{dP_n\left( t\right) }{dt}Q_n\left( t,\lambda \right) .
\label{eq.18}
\end{equation}
Integrating by parts in the last equation we get that 
\begin{equation}
\frac{dP_n\left( \lambda \right) }{d\lambda }=2\int_{-\infty }^{+\infty
}d\mu \left( t\right) Q_n\left( t,\lambda \right) \left( \frac{dV}{dt}-\frac{%
dV}{d\lambda }\right) P_n\left( t\right).  \label{eq.19}
\end{equation}
Now, making use of the Christoffel-Darboux theorem, we conclude that unknown
functions $A_n\left( \lambda \right) $ and $B_n\left( \lambda \right) $ in
Eq. (\ref{eq.17}) are 
\begin{equation}
A_n\left( \lambda \right) =2c_n\int_{-\infty }^{+\infty }\frac{d\mu \left(
t\right) }{t-\lambda }\left( \frac{dV}{dt}-\frac{dV}{d\lambda }\right)
P_n^2\left( t\right) ,  \label{eq.22a}
\end{equation}
\begin{equation}
B_n\left( \lambda \right) =2c_n\int_{-\infty }^{+\infty }\frac{d\mu \left(
t\right) }{t-\lambda }\left( \frac{dV}{dt}-\frac{dV}{d\lambda }\right)
P_n\left( t\right) P_{n-1}\left( t\right) .  \label{eq.22b}
\end{equation}
We also notice the identity that directly follows from Eqs. (\ref{eq.22a}), (%
\ref{eq.22b}), (\ref{eq.06}) and from oddness of $dV/d\lambda $: 
\begin{equation}
B_n\left( \lambda \right) +B_{n-1}\left( \lambda \right) -\frac \lambda
{c_n}A_{n-1}\left( \lambda \right) =-2\frac{dV}{d\lambda }.  \label{eq.222}
\end{equation}

Differentiating Eq. (\ref{eq.17}), making use of the recurrence equation Eq.
(\ref{eq.06}), and bearing in mind relation Eq. (\ref{eq.04a}) between $%
P_n\left( \lambda \right) $ and $\varphi _n\left( \lambda \right) $, one can
obtain an {\it exact} differential equation for the wave-functions of
fictitious fermions, that is valid for arbitrary $n$: 
\begin{equation}
\frac{d^2\varphi _n\left( \lambda \right) }{d\lambda ^2}-{\cal F}_n\left(
\lambda \right) \frac{d\varphi _n\left( \lambda \right) }{d\lambda }+{\cal G}%
_n\left( \lambda \right) \varphi _n\left( \lambda \right) =0,  \label{eq.30}
\end{equation}
where 
\begin{equation}
{\cal F}_n\left( \lambda \right) =\frac 1{A_n}\frac{dA_n}{d\lambda },
\label{eq.31b}
\end{equation}
\begin{eqnarray}
{\cal G}_n\left( \lambda \right) =\frac{dB_n}{d\lambda } &+&\frac{c_n}{%
c_{n-1}}A_nA_{n-1}-B_n\left( B_n+2\frac{dV}{d\lambda }+\frac 1{A_n}\frac{dA_n%
}{d\lambda }\right)  \label{eq.31c} \\
&+&\frac{d^2V}{d\lambda ^2}-\left( \frac{dV}{d\lambda }\right) ^2-\frac
1{A_n}\frac{dA_n}{d\lambda }\frac{dV}{d\lambda }.  \nonumber
\end{eqnarray}

Previously, equation of this type was known in the context of the
random-matrix theory only for GUE, where $V\left( \lambda \right) =\lambda
^2/2$. For such a confinement potential both functions $A_n$ and $B_n$ can
easily be computed from Eqs. (\ref{eq.22a}) and (\ref{eq.22b}), and are
given by $A_n\left( \lambda \right) =2c_n$ and $B_n\left( \lambda \right) =0$%
. Taking into account that for GUE $c_n=\sqrt{n/2}$ we end up with ${\cal F}%
_n\left( \lambda \right) =0$ and ${\cal G}_n\left( \lambda \right)
=2n+1-\lambda ^2$. This allows us to interpret $\varphi _n\left( \lambda
\right) $ as a wave-function of the fermion confined by a parabolic
potential: 
\begin{equation}
\frac{d^2\varphi _n^{\rm{GUE}}\left( \lambda \right) }{d\lambda ^2}+\left(
2n+1-\lambda ^2\right) \varphi _n^{\rm{GUE}}\left( \lambda \right) =0.
\label{eq.444}
\end{equation}
In principle, the effective Schr\"odinger equation Eq. (\ref{eq.30}) applies
to general non-Gaussian random-matrix ensembles as well, although the
explicit calculation of ${\cal F}_n\left( \lambda \right) $ and ${\cal G}%
_n\left( \lambda \right) $ in this situation may be a rather complicated
task. However, significant simplifications arise in the thermodynamic limit $%
n=N\gg 1$.

To proceed with derivation of the asymptotic Schr\"odinger equation, we have
to specify the form of confinement potential $V$ introduced by Eq. (\ref
{eq.02}). Choosing the regular part $v\left( \lambda \right) $ to be an even
function, we set 
\begin{equation}
V^{\left( \alpha \right) }\left( \lambda \right) =\sum_{k=1}^p\frac{d_k}{2k}%
\lambda ^{2k}-\alpha \log \left| \lambda \right| ,  \label{eq.32}
\end{equation}
with $d_p>0$. The signs of the rest $d_k$'s can be arbitrary, allowing for
non-monotonic level confining, but they should lead to an eigenvalue density
supported on a single connected interval $\left( -D_N,+D_N\right) $.
Confinement potential $V^{\left( \alpha \right) }\left( \lambda \right) $
determines its own set of orthogonal polynomials $P_n^{\left( \alpha \right)
}\left( \lambda \right) $, and functions $A_n^{\left( \alpha \right) }$ and $%
B_n^{\left( \alpha \right) }$ which are needed to construct an asymptotic
second-order differential equation for the function $\varphi _N^{\left(
\alpha \right) }\left( \lambda \right) =\left| \lambda \right| ^\alpha
P_N^{\left( \alpha \right) }\left( \lambda \right) \exp \left\{ -v\left(
\lambda \right) \right\} $. [Here upper index $\alpha $ reflects the
presence of the log-singular component in $V^{\left( \alpha \right) }\left(
\lambda \right) $, and the restriction $\alpha >-\frac 12$ takes place due
to normalization Eq. (\ref{eq.05})].

In accordance with Eqs. (\ref{eq.22a}) and (\ref{eq.32}) it is convenient to
represent $A_N^{\left( \alpha \right) }$ in the form 
\begin{equation}
A_N^{\left( \alpha \right) }\left( \lambda \right) =A_{\rm{reg}}^{\left(
N\right) }\left( \lambda \right) +\alpha A_{\rm{sing}}^{\left( N\right)
}\left( \lambda \right) ,  \label{eq.33}
\end{equation}
where 
\begin{equation}
A_{\rm{reg}}^{\left( N\right) }\left( \lambda \right) =2c_N\int_{-\infty
}^{+\infty }\frac{d\mu \left( t\right) }{t-\lambda }\left( P_N^{\left(
\alpha \right) }\left( t\right) \right) ^2\left( \frac{dv}{dt}-\frac{dv}{%
d\lambda }\right) ,  \label{eq.33a}
\end{equation}
\begin{equation}
A_{\rm{sing}}^{\left( N\right) }\left( \lambda \right) =2c_N\int_{-\infty
}^{+\infty }\frac{d\mu \left( t\right) }t\left( P_N^{\left( \alpha \right)
}\left( t\right) \right) ^2.  \label{eq.33b}
\end{equation}
Analogously, Eq. (\ref{eq.22b}) leads to the similar representation 
\begin{equation}
B_N^{\left( \alpha \right) }\left( \lambda \right) =B_{\rm{reg}}^{\left(
N\right) }\left( \lambda \right) +\alpha B_{\rm{sing}}^{\left( N\right)
}\left( \lambda \right) ,  \label{eq.44}
\end{equation}
with 
\begin{equation}
B_{\rm{reg}}^{\left( N\right) }\left( \lambda \right) =2c_N\int_{-\infty
}^{+\infty }\frac{d\mu \left( t\right) }{t-\lambda }P_N^{\left( \alpha
\right) }\left( t\right) P_{N-1}^{\left( \alpha \right) }\left( t\right)
\left( \frac{dv}{dt}-\frac{dv}{d\lambda }\right) ,  \label{eq.44a}
\end{equation}
\begin{equation}
B_{\rm{sing}}^{\left( N\right) }\left( \lambda \right) =\frac{2c_N}\lambda
\int_{-\infty }^{+\infty }\frac{d\mu \left( t\right) }tP_N^{\left( \alpha
\right) }\left( t\right) P_{N-1}^{\left( \alpha \right) }\left( t\right) .
\label{eq.44b}
\end{equation}
In the above formulas $A_{\rm{reg}}^{\left( N\right) }$ and $B_{\rm{reg}%
}^{\left( N\right) }$ result from the regular component of confinement
potential, while $A_{\rm{sing}}^{\left( N\right) }$ and $B_{\rm{sing}%
}^{\left( N\right) }$ are caused by its log-singular part.

First, it is easy to see that $A_{\rm{sing}}^{\left( N\right) }\left(
\lambda \right) \equiv 0$ due to evenness of the measure $d\mu $. Second,
the exact expression for $B_{\rm{sing}}^{\left( N\right) }$ immediately
follows from the recurrence equation Eq. (\ref{eq.06}), whence we get 
\begin{equation}
B_{\rm{sing}}^{\left( N\right) }\left( \lambda \right) =\frac{1-\left(
-1\right) ^N}\lambda .  \label{eq.49}
\end{equation}
Calculation of the regular parts $A_{\rm{reg}}^{\left( N\right) }$ and $B_{%
\rm{reg}}^{\left( N\right) }$ can be done along the lines presented in
Ref. \cite{Kanzieper-1997}. Then, we immediately obtain that $A_N^{\left(
\alpha \right) }\left( \lambda \right) $ is expressed in terms of Dyson's
density $\nu _D\left( \lambda \right) $ as follows: 
\begin{equation}
A_N^{\left( \alpha \right) }\left( \lambda \right) =\frac{\pi \nu _D\left(
\lambda \right) }{\sqrt{1-\left( \lambda /D_N\right) ^2}},  \label{eq.41}
\end{equation}
\begin{equation}
\nu _D\left( \lambda \right) =\frac 2{\pi ^2}{\cal P}\int_0^{D_N}\frac{\xi
d\xi }{\xi ^2-\lambda ^2}\frac{dv}{d\xi }\sqrt{\frac{1-\left( \lambda
/D_N\right) ^2}{1-\left( \xi /D_N\right) ^2}},  \label{eq.42}
\end{equation}
where $D_N=2c_N$ should be identified with the soft edge of the spectrum.
[It is easy to see that a log-singular part of the confinement potential
does not contribute to the Dyson's density, so that in the thermodynamic
limit there are no changes in $D_N$ due to logarithmic singularity of
confinement potential]. Expression for $B_{\rm{reg}}^{\left( N\right) }$
can be obtained by the use of the large$-N$ version of the identity Eq. (\ref
{eq.222}), that yields 
\begin{equation}
B_{\rm{reg}}^{\left( N\right) }\left( \lambda \right) =\frac \lambda
{D_N}A_N^{\left( \alpha \right) }\left( \lambda \right) -\frac{dv}{d\lambda }%
,  \label{eq.50}
\end{equation}
whence 
\begin{equation}
B_N^{\left( \alpha \right) }\left( \lambda \right) =\frac \lambda {D_N}\frac{%
\pi \nu _D\left( \lambda \right) }{\sqrt{1-\left( \lambda /D_N\right) ^2}}-%
\frac{dv}{d\lambda }+\alpha \frac{1-\left( -1\right) ^N}\lambda .
\label{eq.51}
\end{equation}

Now, having asymptotic representations for $A_N^{\left( \alpha \right) }$
and $B_N^{\left( \alpha \right) }$ given by Eqs. (\ref{eq.41}) and (\ref
{eq.51}), and taking into account Eqs. (\ref{eq.30}), (\ref{eq.31b}) and (%
\ref{eq.31c}), it is straightforward to obtain the following remarkable
effective asymptotic Schr\"odinger equation for the wave-functions $\varphi
_N^{\left( \alpha \right) }\left( \lambda \right) =\left| \lambda \right|
^\alpha P_N^{\left( \alpha \right) }\left( \lambda \right) \exp \left\{
-v\left( \lambda \right) \right\} $ of highly excited states $\left( N\gg
1\right) $ of fictitious fermions: 
\begin{eqnarray}
\frac{d^2\varphi _N^{\left( \alpha \right) }}{d\lambda ^2}&-&\left[ \frac
d{d\lambda }\log \left( \frac{\pi \nu _D\left( \lambda \right) }{\sqrt{%
1-\left( \lambda /D_N\right) ^2}}\right) \right] \frac{d\varphi _N^{\left(
\alpha \right) }}{d\lambda }  \label{eq.53} \\
&+&\left[ \pi ^2\nu _D^2\left( \lambda \right) +\frac{\left( -1\right)
^N\alpha -\alpha ^2}{\lambda ^2}\right] \varphi _N^{\left( \alpha \right)
}\left( \lambda \right) =0  \nonumber
\end{eqnarray}
Also, due to Eq. (\ref{eq.17}), one can verify that the wave-functions of
two successive quantum states are connected by the relationship 
\begin{equation}
\frac{d\varphi _N^{\left( \alpha \right) }}{d\lambda }=\frac{\pi \nu
_D\left( \lambda \right) }{\sqrt{1-\left( \lambda /D_N\right) ^2}}\left(
\varphi _{N-1}^{\left( \alpha \right) }\left( \lambda \right) -\frac \lambda
{D_N}\varphi _N^{\left( \alpha \right) }\left( \lambda \right) \right)
+\left( -1\right) ^N\frac \alpha \lambda \varphi _N^{\left( \alpha \right)
}\left( \lambda \right) .  \label{eq.54}
\end{equation}

Equations (\ref{eq.53}) and (\ref{eq.54}) provide a general basis for the
study of eigenvalue correlations in non-Gaussian random-matrix ensembles in
an {\it arbitrary spectral range}. In particular case of GUE, the Dyson's
density of states is the celebrated semicircle, $\nu _D^{\rm{GUE}}\left(
\lambda \right) =\pi ^{-1}\sqrt{D_N^2-\lambda ^2}$ with $D_N=\sqrt{2N}$. The
square-root law for $\nu _D^{\rm{GUE}}\left( \lambda \right) $ immediately
removes the first derivative $d\varphi _N^{\left( \alpha \right) }/d\lambda $
in Eq. (\ref{eq.53}), providing us the possibility to interpret the
fictitious fermions as those confined by a quadratic potential $\left(
\alpha =0\right) $. As far as the semicircle is a distinctive feature of
density of states in GUE only, one will always obtain a first derivative in
the effective Schr\"odinger equation for the non-Gaussian unitary ensembles
of random matrices. Therefore, fictitious non-interacting fermions
associated with non-Gaussian ensembles of random matrices live in a
non-Hermitian quantum mechanics.

An interesting property of these equations is that they do not contain the
regular part of confinement potential explicitly, but only involve the {\it %
Dyson's density} $\nu _D$ (analytically continued on the entire real axis)
and the spectrum end point $D_N$. In contrast, the logarithmic singularity
(that does not affect the Dyson's density) introduces additional singular
terms into Eqs. (\ref{eq.53}) and (\ref{eq.54}), changing significantly the
behavior of the wave-function $\varphi _N^{\left( \alpha \right) }$ near the
origin $\lambda =0$. The influence of the singularity decreases rather
rapidly outward from the origin.

Structure of the effective Schr\"odinger equation leads us to the following
fundamental statements: (i) {\it Eigenvalue correlations are stable with
respect to non-singular deformations of the confinement potential.} (ii) In
the random-matrix ensembles with well-behaved confinement potential {\it the
knowledge of Dyson's density} (that is rather crude one-point
characteristics coinciding with the real density of states only in the
spectrum bulk) {\it is sufficient to determine the genuine density of
states, as well as the }$n-${\it point correlation function, everywhere.}
The latter conclusion is rather unexpected since it considerably reduces the
knowledge required for computing $n-$point correlators.

\section{Local eigenvalue correlations}

Effective Schr\"odinger equation obtained in the preceding Section allows us
to examine in a unified way the eigenvalue correlations in non-Gaussian
ensembles with $U\left( N\right) $ symmetry in different scaling limits. As
we show below, it inevitably leads to the universal Bessel correlations in
the origin scaling limit \cite{Nishigaki-1996,Akemann-1996}, to the
universal sine correlations in the bulk scaling limit \cite
{Brezin-Zee-1993,Eynard-1994,Weidenmuller-1995,FKY-1996}, and to the
universal \rm{G}$-$correlations in the soft-edge scaling limit \cite
{Kanzieper-1997}. Corresponding two-point kernels are given by Eqs. (\ref
{eq.58}), (\ref{eq.59}) and (\ref{eq.0509}), respectively.

\subsection{Origin scaling limit and the universal Bessel law}

Origin scaling limit deals with the region of spectrum close to $\lambda =0$
where confinement potential displays the logarithmic singularity. In the
vicinity of the origin the Dyson's density can be taken as being
approximately a constant, $\nu _D\left( 0\right) =1/\Delta _N\left( 0\right) 
$, where $\Delta _N\left( 0\right) $ is the mean level spacing at the origin
in the absence of the logarithmic deformation of potential $v$. In the
framework of this approximation, Eq. (\ref{eq.53}) reads 
\begin{equation}
\frac{d^2\varphi _N^{\left( \alpha \right) }}{d\lambda ^2}+\left( \frac{\pi
^2}{\Delta _N^2\left( 0\right) }+\frac{\left( -1\right) ^N\alpha -\alpha ^2}{%
\lambda ^2}\right) \varphi _N^{\left( \alpha \right) }\left( \lambda \right)
=0.  \label{eq.55}
\end{equation}
Solution to this equation that remains finite at $\lambda =0$ can be
expressed by means of Bessel functions:

\begin{eqnarray}
\varphi _{2N}^{\left( \alpha \right) }\left( \lambda \right) &=&a\sqrt{%
\lambda }J_{\alpha -\frac 12}\left( \frac{\pi \lambda }{\Delta \left(
0\right) }\right) ,  \label{eq.56a} \\
\varphi _{2N+1}^{\left( \alpha \right) }\left( \lambda \right) &=&b\sqrt{%
\lambda }J_{\alpha +\frac 12}\left( \frac{\pi \lambda }{\Delta \left(
0\right) }\right) ,  \label{eq.56b}
\end{eqnarray}
where $a$ and $b$ are constants to be determined later, and $\Delta \left(
0\right) =\Delta _{2N}\left( 0\right) \approx \Delta _{2N+1}\left( 0\right) $%
. In accordance with Eq. (\ref{eq.08}), the two-point kernel can be written
down as 
\begin{eqnarray}
K_{2N}^{\left( \alpha \right) }\left( \lambda ,\lambda ^{\prime }\right) &=&c%
\frac{\sqrt{\lambda \lambda ^{\prime }}}{\lambda ^{\prime }-\lambda }\left[
J_{\alpha +\frac 12}\left( \frac{\pi \lambda }{\Delta \left( 0\right) }%
\right) J_{\alpha -\frac 12}\left( \frac{\pi \lambda ^{\prime }}{\Delta
\left( 0\right) }\right) \right.  \label{eq.57} \\
&&\left. -J_{\alpha +\frac 12}\left( \frac{\pi \lambda ^{\prime }}{\Delta
\left( 0\right) }\right) J_{\alpha -\frac 12}\left( \frac{\pi \lambda }{%
\Delta \left( 0\right) }\right) \right] ,  \nonumber
\end{eqnarray}
where the unknown factor $c$ can be found from the requirement $%
K_{2N}^{\left( \alpha =0\right) }\left( \lambda ,\lambda \right) =1/\Delta
\left( 0\right) $. This immediately yields us $c=-\pi /\Delta \left(
0\right) $. Defining now the scaled variable $s=\lambda _s/\Delta \left(
0\right) $, we obtain that in the origin scaling limit the two-point kernel $%
K_{\rm{orig}}\left( s,s^{\prime }\right) =\lim_{N\rightarrow \infty
}\left[ K_N^{\left( \alpha \right) }\left( \lambda _s,\lambda _{s^{\prime
}}\right) d\lambda _s/ds\right] $ takes the universal Bessel law 
\begin{equation}
K_{\rm{orig}}\left( s,s^{\prime }\right) =\frac \pi 2\sqrt{ss^{\prime }}%
\frac{J_{\alpha +\frac 12}\left( \pi s\right) J_{\alpha -\frac 12}\left( \pi
s^{\prime }\right) -J_{\alpha -\frac 12}\left( \pi s\right) J_{\alpha +\frac
12}\left( \pi s^{\prime }\right) }{s-s^{\prime }}.  \label{eq.58}
\end{equation}

Formula (\ref{eq.58}) is valid for arbitrary $\alpha >-\frac 12$. Note, that
a recent proof of universality of the Bessel kernel given in Ref. \cite
{Akemann-1996} was based on the Christoffel theorem \cite{Szego-monograph},
that imposed an artificial restriction on parameter $\alpha $ to be only
positive integer.

\subsection{Bulk scaling limit and the universal sine law}

Bulk scaling limit has been explored in a number of works \cite
{Brezin-Zee-1993,Eynard-1994,Weidenmuller-1995,FKY-1996}. It is associated
with a spectrum range where the confinement potential is well behaved (that
is far from the logarithmic singularity $\lambda =0$), and where the density
of states can be taken as being approximately a constant on the scale of a
few levels. In accordance with this definition one has 
\begin{equation}
K_{\rm{bulk}}\left( s,s^{\prime }\right) =\lim_{s,s^{\prime }\rightarrow
\infty }K_{\rm{orig}}\left( s,s^{\prime }\right) ,  \label{eq.588}
\end{equation}
where $s$ and $s^{\prime }$ should remain far enough from the end point $D_N$
of the spectrum support.

Taking this limit in Eq. (\ref{eq.58}), we arrive at the universal sine law 
\begin{equation}
K_{\rm{bulk}}\left( s,s^{\prime }\right) =\frac{\sin \left[ \pi \left(
s-s^{\prime }\right) \right] }{\pi \left( s-s^{\prime }\right) }.
\label{eq.59}
\end{equation}

\subsection{Soft-edge scaling limit and the universal \rm{G}$-$%
multicritical law}

Soft-edge scaling limit is relevant to the tail of eigenvalue support where
crossover occurs from a non-zero density of states to a vanishing one. It is
known \cite{Bowick} that by tuning coefficients $d_k$ which enter the
regular part $v$ of confinement potential [see Eq. (\ref{eq.32})], one can
obtain a bulk (Dyson's) density of states which possesses a singularity of
the type 
\begin{equation}
\nu _D\left( \lambda \right) =\left( 1-\frac{\lambda ^2}{D_N^2}\right)
^{m+1/2}{\cal R}_N\left( \frac \lambda {D_N}\right)  \label{eq.279}
\end{equation}
with the multicritical index $m=0,2,4,$ etc., and ${\cal R}_N$ being a
well-behaved function with ${\cal R}_N\left( \pm 1\right) \neq 0$. [Odd
indices $m$ are inconsistent with our choice that the leading coefficient $%
d_p$, entering the regular component $v\left( \lambda \right) $ of
confinement potential, be positive in order to keep a convergence of
integral for partition function ${\cal Z}_N$ in Eq. (\ref{eq.01})]. Such an $%
m-$th multicriticality can be achieved by many means, and the corresponding
plethora of multicritical potentials $V^{\left( m\right) }$ is given by the
equation 
\begin{equation}
\frac{dV^{\left( m\right) }\left( \lambda \right) }{d\lambda }={\cal P}%
\int_{-D_N}^{+D_N}\frac{dt}{\lambda -t}\left( 1-\frac{t^2}{D_N^2}\right)
^{m+1/2}{\cal R}_N\left( \frac t{D_N}\right) .  \label{eq.278}
\end{equation}
So-called minimal multicritical potentials which correspond to ${\cal R}_N$ $%
=\rm{const}$ can be found in Refs. \cite{Bowick,Migdal}.

Below we demonstrate that as long as multicriticality of order $m$ is
reached, the eigenvalue correlations in the vicinity of the soft edge become
universal, and are independent of the particular potential chosen. The order 
$m$ of the multicriticality is the only parameter which governs spectral
correlations in the soft-edge scaling limit.

Let us move the spectrum origin to its endpoint $D_N$, making the
replacement 
\begin{equation}
\lambda _s=D_N\left[ 1+s\cdot \frac 12\left( \frac 2{\pi D_N{\cal R}_N\left(
1\right) }\right) ^{1/\nu ^{*}}\right] ,  \label{eq.289}
\end{equation}
that defines the $m-$th {\it soft-edge scaling limit} provided $s\ll \left(
D_N{\cal R}_N\left( 1\right) \right) ^{1/\nu ^{*}}\propto N^{1/\nu ^{*}}$,
with 
\begin{equation}
\nu ^{*}=m+\frac 32.  \label{eq.2899}
\end{equation}
It is straightforward to show from Eqs. (\ref{eq.53}) and (\ref{eq.54}) that
in the vicinity of the end point $D_N$ the function $\widehat{\varphi }%
_N\left( s\right) =\varphi _N^{\left( \alpha \right) }\left( \lambda
-D_N\right) $ obeys the universal differential equation 
\begin{equation}
\widehat{\varphi }_N^{\prime \prime }\left( s\right) -\frac{\left( \nu
^{*}-\frac 32\right) }s\widehat{\varphi }_N^{\prime }\left( s\right)
-s^{2\left( \nu ^{*}-1\right) }\widehat{\varphi }_N\left( s\right) =0,
\label{eq.299}
\end{equation}
and that the following relation takes place: 
\begin{equation}
\widehat{\varphi }_{N-1}\left( s\right) =\widehat{\varphi }_N\left( s\right)
+\left( -1\right) ^{\nu ^{*}-\frac 32}\left( \frac 2{\pi D_N{\cal R}_N\left(
1\right) }\right) ^{\frac 1{2\nu ^{*}}}s^{\frac 32-\nu ^{*}}\widehat{\varphi 
}_N^{\prime }\left( s\right) .  \label{eq.309}
\end{equation}

Solution to Eq. (\ref{eq.299}) which decreases at $s\rightarrow +\infty $
(that is at far tails of the density of states) can be represented through
the function 
\begin{eqnarray}
&&\rm{G}\left( s|\nu ^{*}\right) =\frac 1{2\sqrt{\nu ^{*}}}\left[ \sin
\left( \frac \pi {4\nu ^{*}}\right) +\left( -1\right) ^{\nu ^{*}-\frac
32}\right] ^{-1/2}  \label{eq.069} \\
\ \ &&\times \left\{ 
\begin{array}{ll}
s^{\frac 12\left( \nu ^{*}-\frac 12\right) }\left[ I_{-\frac 12\left(
1-\frac 1{2\nu ^{*}}\right) }\left( \frac{s^{\nu ^{*}}}{\nu ^{*}}\right)
-I_{\frac 12\left( 1-\frac 1{2\nu ^{*}}\right) }\left( \frac{s^{\nu ^{*}}}{%
\nu ^{*}}\right) \right] , & s>0, \\ 
\left| s\right| ^{\frac 12\left( \nu ^{*}-\frac 12\right) }\left[ J_{-\frac
12\left( 1-\frac 1{2\nu ^{*}}\right) }\left( \frac{\left| s\right| ^{\nu
^{*}}}{\nu ^{*}}\right) +\left( -1\right) ^{\nu ^{*}-\frac 32}J_{\frac
12\left( 1-\frac 1{2\nu ^{*}}\right) }\left( \frac{\left| s\right| ^{\nu
^{*}}}{\nu ^{*}}\right) \right] , & s<0,
\end{array}
\right.  \nonumber
\end{eqnarray}
[where $J_{\pm \frac 12\left( 1-\frac 1{2\nu ^{*}}\right) }$ and $I_{\pm
\frac 12\left( 1-\frac 1{2\nu ^{*}}\right) }$ are the Bessel functions] as
follows: 
\begin{equation}
\widehat{\varphi }_N\left( s\right) =a\rm{G}\left( s|\nu ^{*}\right) ,
\label{eq.0699}
\end{equation}
where $a$ is an unknown constant. Making use of Eq. (\ref{eq.309}), we
obtain that in the vicinity of the soft edge the two-point kernel is 
\begin{equation}
K_N\left( \lambda _s,\lambda _{s^{\prime }}\right) =b\frac{\rm{G}\left(
s|\nu ^{*}\right) \rm{G}^{\prime }\left( s^{\prime }|\nu ^{*}\right)
\cdot s^{\frac 32-\nu ^{*}}-\rm{G}\left( s^{\prime }|\nu ^{*}\right) 
\rm{G}^{\prime }\left( s|\nu ^{*}\right) \cdot \left( s^{\prime
}\right) ^{\frac 32-\nu ^{*}}}{s-s^{\prime }}\rm{,}  \label{eq.0599}
\end{equation}
where $b$ is an unknown constant again. It can be found by fitting \cite
{Airy-1997} the density of states $K_N\left( \lambda _s,\lambda _s\right) $,
Eq. (\ref{eq.0599}), to the Dyson's density of states $\nu _D\left( \lambda
_s\right) $, Eq. (\ref{eq.279}), near the soft edge provided $1\ll s\ll
N^{1/\nu ^{*}}$. This yields us the value $b=$ $c_N^{-1}\left( \pi c_N{\cal R%
}_N\left( 1\right) \right) ^{1/\nu ^{*}}$. Thus, we obtain that in the $m-$%
th soft-edge scaling limit, Eq. (\ref{eq.289}), the two-point kernel $K_{%
\rm{soft}}^{\left( m\right) }\left( s,s^{\prime }\right)
=\lim_{N\rightarrow \infty }\left[ K_N\left( \lambda _s,\lambda _{s^{\prime
}}\right) d\lambda _s/ds\right] $ satisfies the universal law 
\begin{equation}
K_{\rm{soft}}^{\left( m\right) }\left( s,s^{\prime }\right) =\frac{%
\rm{G}\left( s|\nu ^{*}\right) \rm{G}^{\prime }\left( s^{\prime
}|\nu ^{*}\right) \cdot s^{\frac 32-\nu ^{*}}-\rm{G}\left( s^{\prime
}|\nu ^{*}\right) \rm{G}^{\prime }\left( s|\nu ^{*}\right) \cdot \left(
s^{\prime }\right) ^{\frac 32-\nu ^{*}}}{s-s^{\prime }}\rm{.}
\label{eq.0509}
\end{equation}

These $\rm{G}-$multicritical correlations are universal in the sense
that they do not depend on the details of confinement potential, but only
involve such an ``integral'' characteristic of level confinement as the
index $m$ of the multicriticality. In particular case of $m=0$, that is
inherent in random-matrix ensembles with monotonic confinement potential,
the function $\rm{G}$ coincides with the Airy function, $\rm{G}%
\left( s|\frac 32\right) =\rm{Ai}\left( s\right) $, and the previously
supposed universal Airy correlations \cite{Airy-1997} 
\begin{equation}
K_{\rm{soft}}^{\left( 0\right) }\left( s,s^{\prime }\right) =\frac{%
\rm{Ai} \left( s\right) \rm{Ai} ^{\prime }\left( s^{\prime
}\right) -\rm{Ai} \left( s^{\prime }\right) \rm{Ai} ^{\prime
}\left( s\right) }{s-s^{\prime }}  \label{eq.04}
\end{equation}
are recovered.

It follows from Eq. (\ref{eq.0509}) that the density of states in the same
scaling limit 
\begin{equation}
\nu _{\rm{soft}}^{\left( m\right) }\left( s\right) =\left( \frac d{ds}%
\rm{G} \left( s|\nu ^{*}\right) \right) ^2s^{\frac 32-\nu ^{*}}-\left[ 
\rm{G} \left( s|\nu ^{*}\right) \right] ^2s^{\nu ^{*}-\frac 12}
\label{eq.329}
\end{equation}
is also universal. The large$-\left| s\right| $ behavior of $\nu _{\rm{soft%
}}^{\left( m\right) }$ can be deduced from the known asymptotic expansions
of the Bessel functions: 
\begin{equation}
\nu _{\rm{soft}}^{\left( m\right) }\left( s\right) =\left\{ 
\begin{array}{ll}
\frac{\left| s\right| ^{\nu ^{*}-1}}\pi +\frac{\left( -1\right) ^{\nu
^{*}-\frac 12}}{4\pi \left| s\right| }\cos \left( \frac{2\left| s\right|
^{\nu ^{*}}}{\nu ^{*}}\right) , & s\rightarrow -\infty , \\ 
\frac{\exp \left( -\frac{2s^{\nu ^{*}}}{\nu ^{*}}\right) }{4\pi s}\frac{\cos
^2\left( \frac \pi {4\nu ^{*}}\right) }{\sin \left( \frac \pi {4\nu
^{*}}\right) +\left( -1\right) ^{\nu ^{*}-\frac 32}}, & s\rightarrow +\infty
.
\end{array}
\right.  \label{eq.339}
\end{equation}
Note that the leading order behavior as $s\rightarrow -\infty $ is
consistent with the $\left| s\right| ^{\nu ^{*}-1}$ singularity of the bulk
density of states, Eq. (\ref{eq.279}).

\section{Concluding remarks}

We have presented a general formalism for a treatment of the problem of
eigenvalue correlations in spectra of $U\left( N\right) $ invariant
ensembles of large random matrices with log-singular level confinement. An
important ingredient of our analysis is an effective one-particle
Schr\"odinger equation [see Eqs. (\ref{eq.30}) and (\ref{eq.53})] for
fictitious non-interacting fermions naturally appearing in the determinantal
representation of the joint distribution function of $N$ eigenvalues of
large $N\times N$ Hermitian random matrix. The structure of the asymptotic
equation Eq. (\ref{eq.53}) allowed us to conclude that: (i) Eigenvalue
correlations are stable with respect to non-singular deformations of
confinement potential. (ii) In the random-matrix ensembles with well-behaved
confinement potential the knowledge of Dyson's density (that is rather crude
one-point characteristics coinciding with the real density of states only in
the spectrum bulk) is sufficient to determine the genuine density of states,
as well as the $n-$point correlation function, everywhere. We have also
demonstrated that effective Schr\"odinger equation contains all the
information about eigenvalue correlations in arbitrary spectrum range: the
universal Bessel kernel Eq. (\ref{eq.58}) was found to describe eigenvalue
correlations in the origin scaling limit; the universal sine kernel Eq. (\ref
{eq.59}) was revealed in the bulk scaling limit; finally, we have shown that
the soft-edge scaling limit is described by the novel universal $\rm{G}%
- $multicritical kernel Eq. (\ref{eq.0509}).

\begin{center}
{\bf Acknowledgment}
\end{center}

One of the authors (E. K.) acknowledges the support of the Levy Eshkol
Fellowship from the Ministry of Science of Israel.\newpage\

\end{document}